\documentstyle[12pt]{article}
\font\titlefont=cmbx10 scaled \magstep4

\def\lprox{\mathrel{\raise .3ex\hbox{$<$\kern-
.75em\lower1ex\hbox{$\sim$}}}}

\begin{document}
\input{epsf}

\begin{flushright}
\vspace*{-2cm}
gr-qc/9607003 \\ TUTP-96-2 \\ July 1, 1996 \\ Revised Oct. 2, 1996
\vspace*{2cm}
\end{flushright}

\begin{center}
{\titlefont RESTRICTIONS} \\
\vspace {0.15 in}
{\titlefont ON NEGATIVE ENERGY} \\ 
\vspace {0.15 in}
{\titlefont DENSITY}\\
\vspace {0.15 in}
{\titlefont IN FLAT SPACETIME}\\
\vskip .7in
L.H. Ford\footnote{email: ford@cosmos2.phy.tufts.edu} and 
Thomas A. Roman\footnote{Permanent address: Department of Physics and Earth
Sciences, Central Connecticut State University, New Britain, CT 06050 \\
email: roman@ccsu.ctstateu.edu} \\
\vskip .2in
Institute of Cosmology\\
Department of Physics and Astronomy\\
Tufts University\\
Medford, Massachusetts 02155\\
\end{center}

\begin{abstract}

In a previous paper, a bound on the negative energy density seen by 
an arbitrary inertial observer was derived for the free massless, 
quantized scalar field in four-dimensional Minkowski spacetime. This constraint 
has the form of an uncertainty principle-type limitation on the magnitude and 
duration of the negative energy density. That result was obtained after a 
somewhat 
complicated analysis. The goal of the current paper is to present a much 
simpler method for obtaining such constraints. Similar ``quantum inequality'' 
bounds 
on negative energy density are derived for the electromagnetic field, 
and for the massive scalar field in both two and four-dimensional Minkowski 
spacetime.   
    
\end{abstract}
\newpage

\baselineskip=13pt

\section{Introduction}
\label{sec:intro}
 
In quantum field theory, unlike in classical physics, the energy density may 
be unboundedly negative at a spacetime point. Such situations entail violations 
of all the known classical pointwise energy conditions, such as the weak energy 
condition \cite{HE}. This fact has been known for quite sometime \cite{EGJ}. 
Specific examples include the Casimir effect \cite{C,BM} and squeezed states 
of light 
\cite{WKHW}, both of which have observational support. The theoretical 
prediction of 
black hole evaporation \cite{H75} also involves negative energy densities 
and fluxes in a crucial way. On the other hand, if the laws 
of quantum field theory place no restrictions on negative energy, then it 
might be 
possible to produce gross macroscopic effects such as: violation of the second 
law of thermodynamics \cite{F78,D82} or of cosmic censorship \cite{FR90,FR92}, 
traversable wormholes \cite{MT,MTY}, ``warp drive''\cite{WARP},
and possibly time machines \cite{MTY,AEVERETT}. As a result, much effort has 
been recently directed toward determining what constraints, if any, the laws of 
quantum field theory place on negative energy density. One approach involves 
so-called 
``averaged energy conditions'' (see, for example, \cite{T}-\cite{FW}), 
i.e., averaging the local 
energy conditions over timelike or null geodesics. Another method employs 
``quantum inequalities'' (QI's) \cite{F78,F91}, which are  
constraints on the magnitude and duration of negative energy fluxes and 
densities. 
The current paper is another in a series which is exploring the 
ramifications of 
this approach \cite{FR95}-\cite{MITCHWARP}. (For a more comprehensive 
discussion of the history of these topics, see the introductions of 
Refs.\cite{FR95,FRBH} and the references therein.) 

The QI's have the general form of an inverse relation between an integral
involving the the energy density or flux over a finite time interval and
a power of that interval. More precise 
forms of the inequality were originally derived for negative energy 
fluxes \cite{F91}, and later for negative energy density \cite{FR95,FRBH}. 
This form of QI's involves ``folding'' the stress energy tensor into 
a ``sampling function'', i.e., a peaked function of time whose time 
integral is unity. 
For example, it was shown in Ref.\cite{FR95} that for the free quantized 
massless 
scalar field in four-dimensional Minkowski spacetime,  
\begin{equation}
\hat\rho \equiv {t_0 \over \pi}\, \int_{-\infty}^{\infty}\, 
{{\langle T_{00}\rangle\, dt} 
\over {t^2 + t_0^2}}      \,              
\geq -{3 \over {32 {\pi}^2}{t_0}^4 }  \,,  \label{eq:INTEN}
\end{equation}
for all choices of the sampling time, $t_0$. Here $\langle T_{00}\rangle$ 
is the renormalized expectation value of the energy density evaluated in an 
arbitrary quantum state $|\psi \rangle$, in the frame of an arbitrary 
inertial observer 
whose proper time coordinate is $t_0$. The physical implication of this QI 
is that 
such an observer cannot see unboundedly large negative energy densities 
which persist 
for arbitrarily long periods of time. The QI constraints can be considered 
to be 
midway between the local energy conditions, which are applied at a 
single spacetime 
point, and the averaged energy conditions which are global, in the sense 
that they 
involve averaging over complete or half-complete geodesics. The QI's place 
bounds on the magnitude and duration of the negative energy density in a 
{\it finite neighborhood} of a spacetime point along an observer's worldline.

These inequalities were derived 
for Minkowski spacetime, in the absence of boundaries. However, we recently 
argued \cite{FRWH} that if one is willing to restrict the choice of sampling
time, then the bound should also hold in curved spacetime and/or one with 
boundaries . For example, we proved that the inequality Eq.~(\ref{eq:INTEN}) 
holds in the case of the Casimir effect for sampling times much smaller 
than the distance between the plates. It turns out that this observation has 
some interesting implications for traversable wormholes \cite{FRWH}.
Quantum inequalities in particular curved spacetimes, which reduce to 
Eq.~(\ref{eq:INTEN}) in the short sampling time limit, are given in 
Ref.~\cite{MITCH}.
 
In the original derivation of Eq.~(\ref{eq:INTEN}), we used a rather cumbersome 
expansion of the mode functions of the quantum field in terms of 
spherical waves. 
The goal of the present paper is to present a much more transparent derivation 
of QI bounds, based on a plane wave mode expansion. In so doing, we prove 
new QI constraints on negative energy density for the quantized electromagnetic 
and massive scalar fields. In Sec.~\ref{sec:massive}, we derive a QI bound for 
the massive scalar field in both four and two-dimensional Minkowski spacetime. 
Our earlier result, Eq.~(\ref{eq:INTEN}), is recovered as a special case when 
the mass $m$ goes to zero. A similar bound is obtained for the 
electromagnetic field 
in Sec.~\ref{sec:em}. Our results, and their implications for the existence of 
traversable wormholes, are discussed in Sec.~\ref{sec:summary}. Our metric 
sign convention is $(-,+,+,+)$.

\section{The Massive Scalar Field}
\label{sec:massive}
\subsection{Four-Dimensions}
\label{sec:m4D}  
   In this section we derive a QI-bound on the energy density of a quantized 
uncharged massive scalar field in four-dimensional flat spacetime. The 
wave equation for the field is
\begin{equation}
( \Box - m^2)\phi = 0 \, ,      \label{eq:WE}
\end{equation}
where $\Box \equiv \eta^{\mu \nu}\, \partial_{\mu} \partial_{\nu}$.
We can expand the field operator 
in terms of creation and annihilation
operators as
\begin{equation}
\phi = \sum_{\bf k} ({a_{\bf k}}{f_{\bf k}} + 
a^\dagger_{\bf k} {f_{\bf k}}^\ast). \label{eq:MPHI} 
\end{equation}
Here the mode functions are taken to be
\begin{equation}
f_{\bf k} = {i \over \sqrt{2\omega V}} e^{i({\bf k \cdot x} - \omega t)}, 
\label{eq:MFK}
\end{equation}
where 
\begin{equation}
\omega = \sqrt{ {|{\bf k}|}^2 + m^2} \, ,   \label{eq:MOMEGA}
\end{equation}
$m$ is the rest mass, and $V$ is the normalization volume. The stress tensor 
for the massive scalar field is 
\begin{equation}
T_{\mu \nu} = \phi_{,\mu}\phi_{,\nu} 
-{1\over 2}{\eta_{\mu \nu}(\phi_{,\alpha}\phi^{,\alpha} + m^2 \, 
{\phi}^2) } \, . 
\label{eq:MST}
\end{equation}
The renormalized expectation value of the energy density, in an 
arbitrary quantum state $|\psi \rangle$, is 
\begin{eqnarray}
\langle T_{00} \rangle &=& {{\rm Re}\over {2V}} {\sum_{\bf k',k}} 
{ {(\omega' \omega + {\bf k' \cdot k})} \over {\sqrt{\omega' \omega}} }\, 
\left[\langle a^\dagger_{\bf k'}{a_{\bf k}} \rangle \,
e^{i(\omega'-\omega)t} + \langle {a_{\bf k'}}{a_{\bf k}} \rangle \,
e^{-i(\omega'+\omega)t} \right]          \nonumber\\
&+& {{\rm Re}\over {2V}} {\sum_{\bf k',k}} 
{ {m^2} \over {\sqrt{\omega' \omega}} }\, 
\left[\langle a^\dagger_{\bf k'}{a_{\bf k}} \rangle \,
e^{i(\omega'-\omega)t} - \langle {a_{\bf k'}}{a_{\bf k}} \rangle \,
e^{-i(\omega'+\omega)t} \right]  \,.       \label{eq:MENDEN}        
\end{eqnarray}
Here the energy density is evaluated in the reference frame of an 
inertial observer, at an arbitrary spatial point which we 
choose to be ${ \bf x}=0$. The time coordinate $t$ is the proper time of 
this observer. 
Because of the underlying Lorentz invariance of the field theory, we may 
of course 
choose the frame of any inertial observer in Minkowski spacetime. 

Following Refs. \cite{F91,FR95}, 
we multiply $\langle T_{00}\rangle$ by a peaked function of 
time whose time integral is unity and whose characteristic width 
is $t_0$. A convenient choice of such a function is $t_0 /[\pi(t^2 + t_0^2)]$. 
Define the integrated energy density to be 
\begin{equation}
\hat\rho \equiv {t_0 \over \pi}\, \int_{-\infty}^{\infty}\, 
{{\langle T_{00}\rangle\, dt} 
\over {t^2 + t_0^2}}      \,.              \label{eq:DEFRHOHAT}
\end{equation}
Substitution of Eq.~(\ref{eq:MENDEN}) into Eq.~(\ref{eq:DEFRHOHAT}) and 
performance of the integration yields
\begin{eqnarray}
\hat\rho &=& {{\rm Re}\over {2V}} {\sum_{\bf k',k}} 
{ {(\omega' \omega + {\bf k' \cdot k})} \over {\sqrt{\omega' \omega}} }\, 
\left[\langle a^\dagger_{\bf k'}{a_{\bf k}} \rangle \,
e^{-|\omega'-\omega|t_0} + \langle {a_{\bf k'}}{a_{\bf k}} \rangle \,
e^{-(\omega'+\omega)t_0} \right]          \nonumber\\
&+& {{\rm Re}\over {2V}} {\sum_{\bf k',k}} 
{ {m^2} \over {\sqrt{\omega' \omega}} }\, 
\left[\langle a^\dagger_{\bf k'}{a_{\bf k}} \rangle \,
e^{-|\omega'-\omega|t_0} - \langle {a_{\bf k'}}{a_{\bf k}} \rangle \,
e^{-(\omega'+\omega)t_0} \right]  \,.       \label{eq:MRHOHAT}        
\end{eqnarray}
Now write ${\bf k' \cdot k}= k'_x k_x + k'_y k_y + k'_z k_z$, and apply 
the lemma in 
Appendix B to each of the sums in Eq.~(\ref{eq:MRHOHAT}), e.g.,
\begin{equation}
{\sum_{\bf k',k}} 
{ {k'_x k_x} \over {\sqrt{\omega' \omega}} }\, 
\langle a^\dagger_{\bf k'}{a_{\bf k}} \rangle \,
e^{-|\omega' - \omega|t_0} \geq
{\sum_{\bf k',k}} 
{ {k'_x k_x} \over {\sqrt{\omega' \omega}} }\, 
\langle a^\dagger_{\bf k'}{a_{\bf k}} \rangle \,
e^{-(\omega' + \omega)t_0}       \,.
\end{equation} 
We then obtain
\begin{eqnarray}
\hat\rho &\geq& 
{{\rm Re}\over {2V}} {\sum_{\bf k',k}} 
{ {(\omega' \omega + {\bf k' \cdot k})} \over {\sqrt{\omega' \omega}} }\, 
e^{-(\omega'+\omega)t_0} \,
\left[\langle a^\dagger_{\bf k'}{a_{\bf k}} \rangle \,
+ \langle {a_{\bf k'}}{a_{\bf k}} \rangle  \right]         
\nonumber\\
&+& {{\rm Re}\over {2V}} {\sum_{\bf k',k}} 
{ {m^2} \over {\sqrt{\omega' \omega}} }\,
e^{-(\omega'+\omega)t_0} \, 
\left[\langle a^\dagger_{\bf k'}{a_{\bf k}} \rangle \,
 - \langle {a_{\bf k'}}{a_{\bf k}} \rangle \, \right]  \,.  
\label{eq:TEMPSUM}     
\end{eqnarray}
For the first sum (i.e., the one involving $\omega' \omega$), apply 
the first lemma in 
Appendix A, where we take $h_j = (1/2)(\sqrt{\omega/V})\,e^{-\omega t_0}$. 
Apply the same lemma to each term in the sum which involves 
${\bf k' \cdot k}$, taking 
$h_j = (k_i/{2\sqrt{\omega V}})\,e^{-\omega t_0}$, with $i=x,y,z$. 
For the last sum, 
involving $m^2$, use the second lemma in Appendix A choosing 
$h_j = (m/{2\sqrt{\omega V}})\,e^{-\omega t_0}$. The result is 
\begin{equation}
\hat \rho \geq  -{1 \over {2V} } \, \sum_{\bf k} \, \omega e^{-2 \omega t_0} \,,
\label{eq:RHOSUM}
\end{equation}
where we have used $|{{\bf k}|}^2={\omega}^2-m^2$. 
Now let $V \rightarrow \infty$, in which limit $\sum_{\bf k}\rightarrow 
(V/8 {\pi}^3) \, \int_{-\infty}^{\infty} d^3k$. 
Performing the angular integrations and changing variables we get 
\begin{equation}
\hat \rho \geq -{1 \over {4 {\pi}^2} } \int_m^{\infty} d\omega 
  \sqrt{ {\omega}^2 -m^2 }\, {\omega}^2\,e^{-2 \omega t_0} \,.  
\label{eq:MRHOINT}
\end{equation}

Let $x \equiv 2 \omega t_0,\, y \equiv 2 mt_0$. We may then rewrite 
the integral in Eq.~(\ref{eq:MRHOINT}) as  
\begin{equation}
\hat \rho \geq -{1 \over {64 {\pi}^2}{t_0}^4 } \int_y^{\infty} dx
 \, \sqrt{ x^2 -y^2 }\, x^2\,e^{-x} \,.                \label{eq:MRHOINT/CV}
\end{equation}
If we now let 
\begin{equation}
G(y) \equiv  {1 \over 6} \int_y^{\infty} dx
 \, \sqrt{ x^2 - y^2 }\, x^2\,e^{-x} \,,                \label{eq:DEFG}
\end{equation}
then our QI-bound may be written as 
\begin{equation}
\hat \rho \geq -{3 \over {32 {\pi}^2}{t_0}^4 } \, G(y) \,,       
\label{eq:MQI/4D}
\end{equation}
for all $t_0$. 
A plot of $G(y)$ versus $y$ is given in Fig. 1. 
\begin{figure}
\epsfysize=10cm\epsffile{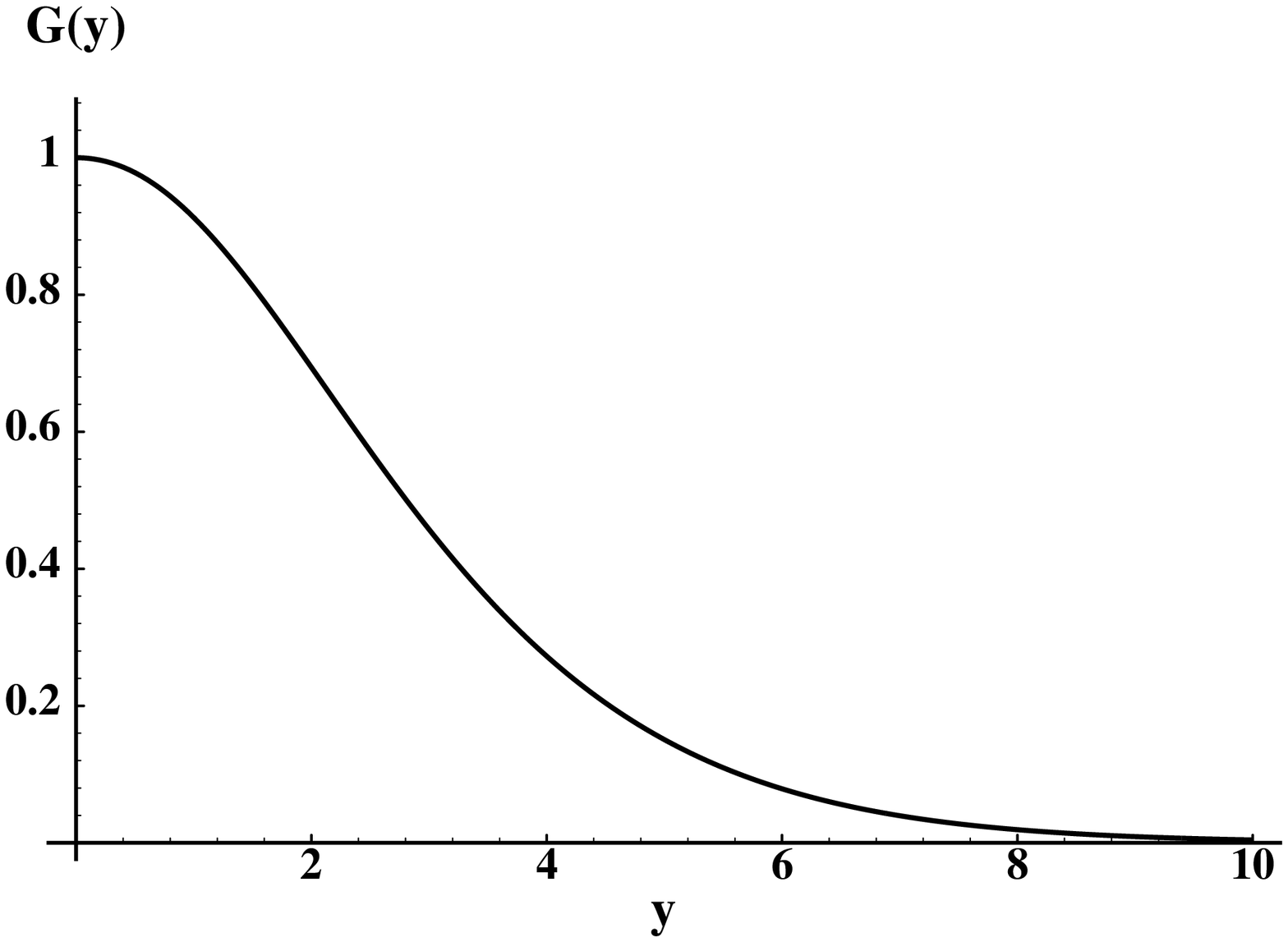}
\label{Figure 1}
\begin{caption}[]

\qquad \qquad \qquad \qquad \qquad \qquad The graph of $G(y)$ versus $y$.

\end{caption}
\end{figure}\par
We see that as $m \rightarrow 0$ 
(for fixed $t_0$), $y \rightarrow 0$ and $G(y) \rightarrow 1$, hence in 
this case our bound reduces to 
\begin{equation}
\hat \rho \geq -{3 \over {32 {\pi}^2}{t_0}^4 }  \,,       \label{eq:QI}
\end{equation}
for all $t_0$, which is the QI for the massless scalar field - originally 
obtained in 
\cite{FR95} by a much more complicated method. 
As $m \rightarrow \infty$ (for fixed $t_0$), $y \rightarrow \infty$, 
and we see from the graph that $G(y) \rightarrow 0$. 
Note that our bound becomes more stringent as $m$ increases, 
and hence it becomes increasingly difficult to produce large negative 
energy densities. 
This result is not surprising, since now one has to overcome the positive 
rest mass energy
of the field quanta. For $t_0 \gg 1/m$ (with fixed $m$), corresponding 
to sampling 
times much larger than the Compton wavelength of the particle, we also have 
$y \rightarrow \infty$ and $G(y) \rightarrow 0$. Note that, due to the factor 
of $G(y)$, the right-hand side of Eq.~(\ref{eq:MQI/4D}) vanishes more rapidly 
than the right-hand side of Eq.~(\ref{eq:QI}) in the $t_0 \rightarrow 
\infty$ limit. This can be interpreted as showing that negative energy due 
to a massive scalar field must be more highly localized than that due to a 
massless field.  

\subsection{Two-Dimensions}
\label{sec:m2D}
The derivation of a QI bound on the energy density of a massive scalar field in 
two-dimensional spacetime is very similar to the discussion given in 
the previous section 
for the four-dimensional case. One follows the same steps as in that treatment, 
but replacing the normalization volume $V$ with a periodicity length $L$, and 
${\bf k}$ and ${\bf x}$ by $k$ and $x$, respectively. This procedure yields 
\begin{equation}
\hat \rho \geq  -{1 \over {2L} } \, \sum_{k} \, \omega e^{-2 \omega t_0} \,,
\label{eq:RHOSUM/2D}
\end{equation}
where we have used $k^2={\omega}^2-m^2$. It is at this stage that the 2D 
analysis differs from the 4D case. Now let $L \rightarrow \infty$, in which 
limit $\sum_{k}\rightarrow 
(L/{2\pi}) \, \int_{-\infty}^{\infty} dk$.  
Rewriting the integral in terms of $\omega$, one obtains  
\begin{equation}
\hat \rho \geq -{1 \over {2\pi} } \int_m^{\infty} \,
{ {{\omega}^2\,e^{-2 \omega t_0}} \over   \sqrt{ {\omega}^2 -m^2 } }
\, d\omega \,.  \label{eq:MRHOINT/2D}
\end{equation}

Let 
\begin{equation}
I  \equiv \int_m^{\infty} \,
{ {{\omega}^2\,e^{-2 \omega t_0}} \over   \sqrt{ {\omega}^2 -m^2 } }
\, d\omega \,.           \label{eq:DEFI}  
\end{equation}
This expression can be rewritten \cite{GR3.365.2} in terms of modified 
Bessel functions of the second kind, $K_{\nu}(y)$, as follows 
\begin{equation}
I = - m^2 \, K'_1 (2 m t_0) \,,   \label{eq:I2} 
\end{equation}
where the prime denotes differentiation with respect to the argument.
We now employ the relation \cite{GR8.486.11} 
\begin{equation}
K'_1 (y) = - {1 \over 2} \, [K_0(y) + K_2(y)] \,,   \label{eq:RECREL}
\end{equation}
where again $y \equiv 2mt_0$. The integral $I$ can then be written as 
\begin{eqnarray}
I &=& {1 \over {8 {t_0}^2} } \, y^2 \, [K_0(y) + K_2(y)]  \nonumber\\
&=& {1 \over {4 {t_0}^2} } \, F(y) \,.     \label{eq:I/F}
\end{eqnarray}
Thus our QI-bound may be written as 
\begin{equation}
\hat \rho \geq -{1 \over {8 \pi {t_0}^2} } \, F(y) \,,       \label{eq:MQI/2D}
\end{equation}
for all $t_0$. 
The function $F(y)$ is plotted against $y$ in Fig. 2.
\begin{figure}
\epsfysize=10cm\epsffile{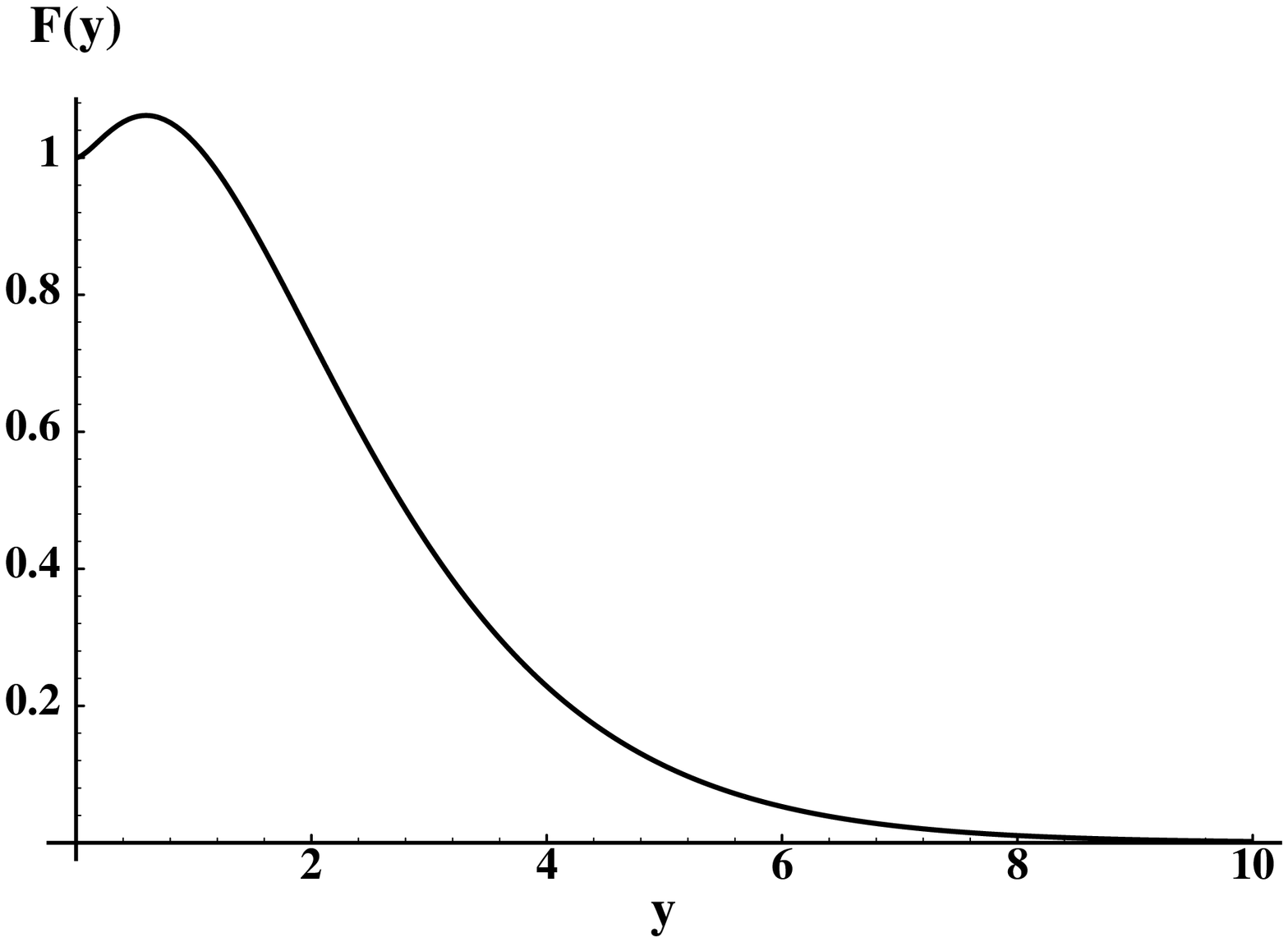}
\begin{caption}[]

\qquad \qquad \qquad \qquad \qquad \qquad The graph of $F(y)$ versus $y$.

\label{Figure 2}  
\end{caption}
\end{figure}\par
 As $m \rightarrow 0$ (for 
fixed $t_0$), $F(y) \rightarrow 1$, and our inequality becomes 
\begin{equation}
\hat \rho \geq -{1 \over {8 \pi {t_0}^2 } } \,,       \label{eq:QI/2D}
\end{equation}
for all $t_0$, which is the QI for the massless scalar field 
in 2D \cite{FR95}. For large $y$, 
the behavior of the graph is qualitatively similar to the 4D case. However, for 
$y$ in the approximate range $0 \leq y \leq 1.2$, there appears to be a 
small peak 
in the value of $F(y)$. In the case of flat spacetime, this seems to be an 
artifact of 2D, since no corresponding 
peak occurs in the 4D case \cite{MITCH/CBUMP}. Rather than indicating 
that one could 
actually supercede the $m=0$ bound in the 2D massive case for certain 
values of $y$, this 
result may perhaps imply that the bound given in Eq.~(\ref{eq:MQI/2D}) 
is not optimal over this range. In either case, the relative height of the peak 
above the value of $F(y)=1$ (corresponding to the massless case) is too 
small to produce 
a dramatic change in the QI bound. As we argued in Ref.\cite{FRWH}, the 
constant on the right-hand side of the inequality would typically have to 
be larger by many orders of magnitude to result in large macroscopic effects.

\section{The Electromagnetic Field}
\label{sec:em}

   In this section, we derive a QI-bound for the energy density of the 
free quantized 
electromagnetic field in 4D Minkowski spacetime. The absence of source 
charges and the 
imposition of the Coulomb gauge condition imply 
\begin{eqnarray}
&{}& A^0  = \Phi = 0  \,,  \nonumber\\ 
&{}& {\bf \nabla  \cdot  A} = 0  \,,      \label{eq:GAUGE}
\end{eqnarray}
where ${\bf A}$ is the vector potential, and $\Phi$ is the scalar potential. 
The wave equation is 
\begin{equation}
{\bf \Box  A} =0  \,.       \label{eq:EMWAVEQ}
\end{equation}
The vector potential can be expanded in terms of creation and 
annihilation operators as
\begin{equation}
{\bf A} ({\bf x},t) = \sum_{{\bf k}\lambda} ({a_{{\bf k} \lambda}} 
\hat e_{{\bf k} \lambda} {f_{\bf k}} + 
a^\dagger_{{\bf k} \lambda} \hat e_{{\bf k} \lambda} {f_{\bf k}}^\ast)\,, 
\label{eq:A} 
\end{equation}
where the $\hat e_{{\bf k} \lambda}$ are unit linear polarization vectors 
and thus 
$\lambda =1,2$. Here the sum $\sum_{{\bf k}\lambda} =
\sum_{\bf k}\,\sum_{\lambda}$. The mode functions $f_{\bf k}$ are 
again given by 
\begin{equation}
f_{\bf k} = {i \over \sqrt{2\omega V}} e^{i({\bf k \cdot x} - \omega t)}, 
\label{eq:EMFK}
\end{equation}
i.e., we assume periodic boundary conditions in a box with 
normalization volume $V$. We will adopt the convention that one of 
the polarization 
vectors changes direction as ${\bf k} \rightarrow -{\bf k}$, but the other 
does not. 
That is, we have 
\begin{eqnarray}
\hat e_{{\bf k} 1} {\bf \times} \hat e_{{\bf k} 2} &=& {\bf {\hat k}} \,,
           \nonumber\\
\hat e_{-{\bf k} 1} &=& -\hat e_{{\bf k}1} \,, \nonumber\\
\hat e_{-{\bf k} 2} &=&  \hat e_{{\bf k} 2}  \,,  \label{eq:DEFe's}
\end{eqnarray} 
so that 
$\hat e_{-{\bf k} 1} {\bf \times} \hat e_{-{\bf k} 2} =  -{\bf {\hat k}}$.
The commutation relations for the creation and annihilation operators are 
\begin{equation}
\quad \qquad [a_{{\bf k} \lambda}, a^\dagger_{{\bf k'} \lambda'}] 
= \delta_{ {\bf k} {\bf k'} } \, \delta_{\lambda \lambda'} \,, \label{eq:CR1}
\end{equation}
\begin{equation}
[a_{{\bf k} \lambda}, a_{{\bf k'} \lambda'}] = 0 \,, \label{eq:CR2}
\end{equation}
\begin{equation}
[a^\dagger_{{\bf k} \lambda }, a^\dagger_{{\bf k'} \lambda'}] = 0 \,. 
\label{eq:CR}
\end{equation}

With the gauge choice made in Eq.~(\ref{eq:GAUGE}), the electric field 
operator becomes 
\begin{equation}
{\bf E} = -\dot {\bf A}
= \sum_{{\bf k}\lambda} \, i \omega \,({a_{{\bf k} \lambda}} 
\hat e_{{\bf k} \lambda} {f_{\bf k}} - 
a^\dagger_{{\bf k} \lambda} \hat e_{{\bf k} \lambda} {f_{\bf k}}^\ast) \,.
\label{eq:EOP}
\end{equation}
Similarly the magnetic field operator is given by 
\begin{equation}
{\bf B} =  {\bf \nabla \times A}
= \sum_{{\bf k}\lambda} \, i \omega \,({a_{{\bf k} \lambda}} 
\hat b_{{\bf k} \lambda} {f_{\bf k}} - 
a^\dagger_{{\bf k} \lambda} \hat b_{{\bf k} \lambda} {f_{\bf k}}^\ast) \,.
\label{eq:BOP}
\end{equation}
Here we define 
\begin{equation}
\hat b_{{\bf k} \lambda}={\bf \hat k } {\bf \times} 
\hat e_{{\bf k} \lambda} \,,  \label{eq:bDEF}
\end{equation} 
so that 
\begin{equation}
\hat b_{{\bf k} 1} = {\bf {\hat k}} {\bf \times}  \hat e_{{\bf k} 1} 
= \hat e_{{\bf k} 2} \,,     \label{eq:b1}
\end{equation}
\begin{equation}
\quad \hat b_{{\bf k} 2} = {\bf {\hat k}} {\bf \times}  \hat e_{{\bf k} 2} 
= -\hat e_{{\bf k} 1} \,,     \label{eq:b2}
\end{equation}
and therefore 
\begin{equation}
\hat b_{-{\bf k} 1} = -{\bf {\hat k}} {\bf \times}  \hat e_{-{\bf k} 1} 
= \hat e_{{\bf k} 2}= \hat b_{{\bf k} 1}  \,,     \label{eq:b(-1)}
\end{equation}
and 
\begin{equation}
\quad \hat b_{-{\bf k} 2} = -{\bf {\hat k}} {\bf \times}  \hat e_{-{\bf k} 2} 
= \hat e_{{\bf k} 1}= - \hat b_{{\bf k} 2}  \,.     \label{eq:b(-2)}
\end{equation}

The stress energy tensor for the electromagnetic field is 
\begin{equation}
T_{00} = {1 \over 2} (E^2 + B^2) \,.  \label{eq:CSTEM}
\end{equation}
If we compute the renormalized expectation value $\langle T_{00} \rangle$, 
in an 
arbitrary quantum state $|\psi \rangle$, evaluated at ${\bf x}=0$, and fold this 
into our 
sampling function we obtain 
\begin{eqnarray}
\hat\rho &\equiv& {t_0 \over \pi}\, \int_{-\infty}^{\infty}\, 
{{\langle T_{00}\rangle\, dt} 
\over {t^2 + t_0^2}}     \nonumber\\
&=& {{\rm Re}\over {2V}} \sum_{ {\bf k,k'} \atop {\lambda, \lambda'} }  
\sqrt{\omega' \omega} \,
[(\hat e_{{\bf k} \lambda} \cdot \hat e_{{\bf k'} \lambda'}) \, + 
\, (\hat b_{{\bf k} \lambda} \cdot \hat b_{{\bf k'} \lambda'}) ] \, 
\nonumber\\
&\qquad&  \times \left[\langle a^\dagger_{ {\bf k} \lambda}  
{a_{ {\bf k'} \lambda' }} \rangle \, e^{-|\omega-\omega'|t_0} + 
\langle {a_{ {\bf k} \lambda } } {a_{ {\bf k'} \lambda' } } \rangle \,
e^{-(\omega+\omega')t_0} \right]   \,.       \label{eq:EMRHOHAT}        
\end{eqnarray}
Now write $\hat e_{{\bf k} \lambda} \cdot \hat e_{{\bf k'} \lambda'} =
{({\hat e}_{ {\bf k} \lambda}) }_x \, {({\hat e}_{{\bf k'} \lambda'})}_x \,
+ \,{({\hat e}_{{\bf k} \lambda}) }_y \, {({\hat e}_{{\bf k'} \lambda'})}_y \,
+ \,{({\hat e}_{{\bf k} \lambda}) }_z \, {({\hat e}_{{\bf k'} \lambda'})}_z $,
and 
similarly for the dot product involving the $\hat b$ vectors. Expanding 
Eq.~(\ref{eq:EMRHOHAT}) in this fashion, one gets six terms. To each of these 
terms we apply the lemma in Appendix B, e.g.,
\begin{eqnarray}
&& \!\!\!\!\!\!\!\!\! \sum_{ {\bf k,k'} \atop {\lambda,\lambda'} }  \, 
\sqrt{\omega \omega'}\,
{({\hat e}_{{\bf k} \lambda}) }_x \, {({\hat e}_{{\bf k'} \lambda'})}_x \,
{\langle a^\dagger_{{\bf k} \lambda} {a}_{{\bf k'} \lambda'} \rangle} \,
e^{- {|\omega -\omega'|} t_0}     \nonumber\\
&& \!\!\!\!\!\!\! \geq \sum_{ {\bf k,k'} \atop {\lambda,\lambda'} } 
\, \sqrt{\omega \omega'}\,
{({\hat e}_{{\bf k} \lambda}) }_x \, {({\hat e}_{{\bf k'} \lambda'})}_x \,
{\langle a^\dagger_{{\bf k} \lambda} \, {a}_{{\bf k'} \lambda'} \rangle} \,
e^{-(\omega +\omega') t_0} \,.       \label{eq:SUMINQ}
\end{eqnarray} 
Recombining the terms, we then obtain
\begin{eqnarray}
\hat \rho \!\! &\geq& \!\!
{{\rm Re}\over {2V}} \sum_{ {\bf k,k'} \atop {\lambda, \lambda'} } 
\sqrt{\omega' \omega} \,
[(\hat e_{{\bf k} \lambda} \cdot \hat e_{{\bf k'} \lambda'}) \, + 
\, (\hat b_{{\bf k} \lambda} \cdot \hat b_{{\bf k'} \lambda'}) ] \, 
\nonumber\\
&\qquad& \times \left[\langle a^\dagger_{ {\bf k} \lambda}  
{a_{ {\bf k'} \lambda' }} \rangle \, + 
\langle {a_{ {\bf k} \lambda } } {a_{ {\bf k'} \lambda' } } \rangle  \right]\,
e^{-(\omega+\omega')t_0}  \,.       \label{eq:EMRHOHAT2}
\end{eqnarray}
An analogous separation of the dot products in Eq.~(\ref{eq:EMRHOHAT2}) 
into components, repeated application of the first lemma in 
Appendix A, and a recombination of terms results in 
\begin{equation}
\hat \rho \geq  -{1 \over {2V} } \, \sum_{ {\bf k}\lambda}
\, \omega e^{-2 \omega t_0} \,,
\label{eq:EMRHOSUM1}
\end{equation}
where we have used $h_j =(1/2)\,{({\hat e}_{{\bf k} \lambda}) }_i \, \times \,
\sqrt{\omega/V} \, e^{-\omega t_0}$ and 
$h_j = (1/2)\,{({\hat b}_{{\bf k} \lambda}) }_i \, \times  
\sqrt{\omega/V} \,
e^{-\omega t_0}$, with $i=x,y,z$.  Evaluation of the sum 
over $\lambda$ gives 
\begin{equation}
\hat \rho \geq  -{1 \over V} \, \sum_{\bf k}
\, \omega e^{-2 \omega t_0} \,.
\label{eq:EMRHOSUM2}
\end{equation}
Now let $V \rightarrow \infty$, so 
\begin{equation}
\hat\rho \geq -{1 \over {2{\pi}^2}}\,\int_{0}^{\infty}\, d{\omega}\,
{\omega}^{3}\,e^{-2 \omega t_0}\,.           \label{eq:EMRHOHAT-INT}
\end{equation}
An evaluation of the integral gives us our desired result:
\begin{equation}
\hat\rho \geq -{3\over {16 {\pi}^2 {t_0}^4}}\,,  \label{eq:EMHATRHO/QI}
\end{equation}
for all $t_0$. Comparison with the QI-bound for the massless scalar field, 
Eq.~(\ref{eq:QI}), shows that the right-hand side of the bound in 
the electromagnetic 
field case is smaller (i.e., more negative) by a factor of $2$. This is 
just what 
one would expect due to the two polarization degrees of freedom.

\section{Conclusions}
\label{sec:summary}
 In the current paper, we have derived ``quantum inequality'' (QI) 
bounds on negative energy density for the quantized 
(uncharged) massive scalar and electromagnetic fields in 
Minkowski spacetime. The bounds take the form of uncertainty principle-type 
inequalities on the magnitude and duration of the negative energy density 
seen by 
an arbitrary inertial observer. (Note however that the energy-time uncertainty 
principle is {\it not} used as input to derive the QI bounds.) 
Originally we had derived such a bound for the 
massless scalar field in both two and four-dimensional Minkowski spacetime 
\cite{FR95}. Our earlier four-dimensional derivation was performed using a 
rather complicated expansion of the mode functions in spherical waves. 
The present 
treatment is considerably simpler, as the analysis is done in a plane wave 
mode representation. We recover our previous results for the massless scalar 
field as the $m=0$ case of the massive scalar field. For the massive 
scalar field, 
we find that in general it is more difficult to obtain sustained 
negative energy 
densities than in the massless case. This is not surprising, as now one must 
overcome the positive rest mass energy. In the case of the electromagnetic 
field, the right-hand side of the bound is slightly weaker, i.e., by a factor 
of $2$, than in the massless scalar field case. This is also to be expected, 
given the two polarization degrees of freedom in the former case. 
For all of our QI bounds on energy density, in the infinite sampling time 
limit we sample over the entire timelike geodesic of 
the observer and obtain the averaged weak energy condition (AWEC):
\begin{equation}
\int_{-\infty}^{\infty} \, T_{\mu \nu} u^{\mu} u^{\nu} \, d\tau \,
\geq 0 \,,
\label{eq:AWEC}
\end{equation}
where $u^{\mu}$ is the unit tangent vector to the geodesic and $\tau$ is 
the observer's proper time. Hence, in Minkowski spacetime, the AWEC can be 
derived from the QI bound. 

Note that in Minkowski spacetime, our bounds hold for an arbitrary choice 
of the 
sampling time, $t_0$. Recently we argued that such bounds should also hold in a 
curved spacetime and/or one with boundaries, if the sampling time is restricted 
to be much smaller than the smallest local radius of curvature and/or 
the distance 
to any boundaries in the spacetime \cite{FRWH}. An explicit example 
of the validity of this assumption has been given in Ref. \cite{MITCH}. 
There a QI is derived in the static closed and open Robertson-Walker
universe models. For a choice of sampling time which is small compared to the 
local radius of curvature, the QI reduces to our flat spacetime bound. 

The application of our bound to Morris-Thorne wormhole spacetimes \cite{FRWH}, 
in the small sampling time limit, implied that typically either a wormhole 
must be only slightly larger than the 
Planck size, or that the negative energy density must be concentrated in an 
extraordinarily thin band near the throat. As an example, in one case for a 
wormhole with a throat radius of $1\,{\rm m}$, the negative energy 
density must be 
concentrated in a band {\it no thicker} than about a millionth of a proton 
radius. Even for a wormhole with a throat radius the size of a galaxy, the band 
of negative energy must be no thicker than about $10$ proton radii (for this 
particular type of wormhole) \cite{COMMENT:MITCHWARP}. That analysis 
assumed that the stress energy 
tensor which maintained the wormhole consisted of quantized massless 
scalar fields. 
More specifically, the $\langle T_{\mu \nu} \rangle$ which generates the 
wormhole geometry was assumed to be the expectation value of the stress energy 
tensor operator of the massless scalar field in a suitable quantum state. 
The results of the present paper indicate that the same restrictions on 
the wormhole length scales will also apply 
if one tries to make the stress energy of the wormhole out of electromagnetic 
or massive scalar fields. One might hope that the constraints imposed by these 
bounds could be circumvented by the superposition of many fields, each 
of which individually satisfies a QI bound. However, we showed \cite{FRWH} 
that in practice to achieve significant macroscopic effects, one needs either 
on the order of $10^{62}$ fields or a few fields for which the numerical 
constant on the right-hand side of the QI bound is many orders of magnitude 
larger than in the cases examined to date. Neither of these possibilities 
seem very likely. It therefore appears probable that Nature will always prevent 
us from producing gross macroscopic effects with negative energy.
 
\vskip 0.2 in
\centerline{\bf Acknowledgements}
We would like to thank Michael Pfenning for useful discussions and for help 
with the graphics. TAR would like to thank the members of the Tufts 
Institute of Cosmology for their warm hospitality while this work was 
being done. 
This research was supported in part by NSF Grant No. PHY-9507351 and by a 
CCSU/AAUP faculty research grant.

\appendix
\section*{Appendix A}
\label{sec:appendixA}
\setcounter{equation}{0}
\renewcommand{\theequation}{A\arabic{equation}}

In this appendix we will establish two lemmas on sums of the expectation 
values of products of creation and annihilation operators. The discussion 
is a slight generalization of the argument presented in the Appendix of 
Ref.\cite{FR92}. The idea for this method of proof was originally suggested 
to us by Flanagan \cite{FR92}. The first lemma was proven by a more 
complicated argument in Appendix A of Ref.\cite{F91}.
\vskip 0.2in
\noindent {\bf Lemma 1}: 
\vskip 0.1in

Define the operator $\hat O_j$ by
\begin{equation}
\hat O_j = \sum_j h_j (a_j + a^\dagger_j) \,,    \label{eq:DEFO}
\end{equation} 
where $j$ is a generalized mode label, and the $h_j$'s are assumed to be real. 
Then it is easily seen that $\hat O_j$ is a Hermitian operator. We may form the 
expectation value
\begin{equation}
\langle {\hat O}^\dagger_j \hat O_j \rangle \geq 0 \,.  \label{eq:OdagO}
\end{equation}
This follows from the fact that the left-hand side is simply the norm of 
the state 
vector $|\Psi \rangle = \hat O_j |\psi \rangle$, where $|\psi \rangle$ 
is the quantum state in which the expectation value is taken. 
We may expand this expression as  
\begin{equation}
\langle {\hat O}^\dagger_j \hat O_j \rangle \, 
= \sum_{j,j'} \, h_{j'} h_j [\langle a^\dagger_{j'} a_j \rangle \, + \,
\langle a^\dagger_j a_{j'} \rangle \, + \, \langle a_{j'} a_j \rangle \, + \,
\langle a^\dagger_{j'} a^\dagger_j \rangle ]\, + \, \sum_j \, {(h_j)}^2  \,.
\label{eq:OSUM1}
\end{equation}
Use the fact that $\langle a^\dagger_{j'} a_j \rangle = 
{\langle a^\dagger_j a_{j'} \rangle}^\ast$ to write 
\begin{equation}
\langle {\hat O}^\dagger_j \hat O_j \rangle \, 
= 2 \,{\rm Re} \, \sum_{j,j'} \, h_{j'} h_j [\langle a^\dagger_{j'} a_j \rangle 
\, + \, \langle a_{j'} a_j \rangle ] \, + \, \sum_j \, {(h_j)}^2  \,.
\label{eq:OSUM2}
\end{equation}
It follows from Eq.~(\ref{eq:OdagO}) that   
\begin{equation}
2 \,{\rm Re} \, \sum_{j,j'} \, h_{j'} h_j [\langle a^\dagger_{j'} a_j \rangle 
\, + \, \langle a_{j'} a_j \rangle ] \, \geq \, - \sum_j \, {(h_j)}^2  \,.
\label{eq:L1}
\end{equation}

\vskip 0.2in
\noindent {\bf Lemma 2}: 
\vskip 0.1in
Define the operator $\hat P_j$ by
\begin{equation}
\hat P_j = i\,\sum_j h_j (a^\dagger_j - a_j) \,,    \label{eq:DEFP}
\end{equation} 
where $j$ is a generalized mode label, and the $h_j$'s are again assumed 
to be real. Then the operator $\hat P_j$ is also Hermitian. Utilizing the 
same chain of reasoning used to establish Lemma 1, one can show that 
\begin{equation}
2 \,{\rm Re} \, \sum_{j,j'} \, h_{j'} h_j [\langle a^\dagger_{j'} a_j \rangle 
\, - \, \langle a_{j'} a_j \rangle ] \, \geq \, - \sum_j \, {(h_j)}^2  \,.
\label{eq:L2}
\end{equation}

\appendix
\section*{Appendix B}
\setcounter{equation}{0}
\renewcommand{\theequation}{B\arabic{equation}}

In this appendix, we wish to prove a lemma which generalizes the inequality
proven in Appendix B of Ref. \cite{F91}.  Consider the sum
\begin{equation}
S_m \equiv \sum_{j j'} p^*_j \, p_{j'}\, \langle a^\dagger_j a_{j'}\rangle
           \, {\rm e}^{-|\omega_j -\omega_{j'}| t_0} \,. \label{eq:S}
\end{equation}
Here the summation is over a finite set of modes with frequencies 
$\omega_j$, with $\omega_m = {\rm max}\{\omega_j \}$, and
$p_j$ is an arbitrary complex function of the mode label $j$. 
(In Ref. \cite{F91} , it was assumed that the $\omega_j \propto j$.
That assumption is not needed here.)
We first note that $S_m$ is real as a consequence of the fact 
that $\langle a^\dagger_j a_{j'}\rangle^* = 
\langle a^\dagger_{j'} a_{j}\rangle$. Next we define the sum
\begin{equation}
\tilde S_m \equiv 
  \sum_{j j'} p^*_j \, p_{j'}\, \langle a^\dagger_j a_{j'}\rangle
           \, {\rm e}^{-(\omega_j +\omega_{j'}) t_0} \,,  
                                              \label{eq:S_tilde}
\end{equation}
which is also real. We wish to prove that
\begin{equation}
S_m \geq  \tilde S_m   \,.
\end{equation}
First note that
\begin{equation}
 \tilde S_m \geq 0   \,.
\end{equation}
This follows from the fact that any sum of the form 
\begin{equation}
\sum_{j j'} p^*_j \, p_{j'}\, \langle a^\dagger_j a_{j'}\rangle
\end{equation}
is non-negative as a consequence of its being the norm of the state
vector
\begin{equation}
|\Psi \rangle = \sum_{j} p_j\, a_j|\psi \rangle \,
\end{equation}
where $|\psi \rangle$ is the quantum state in which the expectation
values in Eqs.~(\ref{eq:S}) and (\ref{eq:S_tilde}) are to be taken.

Let us assume that the eigenfrequencies are discrete, as will be the case
with a finite quantization volume, and that the lowest frequency is greater
than zero. We can then order the modes in increasing frequency so that
$0 < \omega_0 \leq \omega_1 \leq \cdots \leq \omega_m$. In general, there
can be several modes with the same frequency. Let $\ell$ label the distinct
frequencies. and let $n(\ell)$ label the various modes all having frequency
$\omega_\ell$. Then $\sum_j = \sum_\ell \sum_{n(\ell)}$. 

Define
\begin{equation}
A_{jj'} =  p^*_j \, p_{j'}\, \langle a^\dagger_j a_{j'}\rangle
           \, {\rm e}^{-|\omega_j -\omega_{j'}| t_0} \,, 
\end{equation}
and
\begin{equation}
B_{jj'} =  p^*_j \, p_{j'}\, \langle a^\dagger_j a_{j'}\rangle
           \, {\rm e}^{-(\omega_j +\omega_{j'}) t_0} \,, \label{eq:B}
\end{equation}
so that
\begin{equation}
S_m -  \tilde S_m = \sum_{jj'} (A_{jj'} - B_{jj'})   \,.
\end{equation}
Note that
\begin{equation}
\omega_j +\omega_{j'} - |\omega_j -\omega_{j'}| = 2\, \Omega_{jj'} \,,
\end{equation}
where $\Omega_{jj'}$ is the lesser of $\omega_j$ and $\omega_{j'}$.
We can then write
\begin{equation}
A_{jj'} -B_{jj'} =  p^*_j \, p_{j'}\, \langle a^\dagger_j a_{j'}\rangle
\bigl({\rm e}^{2 \Omega_{jj'}\, t_0} -1 \bigr)\, 
                                   {\rm e}^{-(\omega_j +\omega_{j'}) t_0} \,. 
                   \label{eq:AB}
\end{equation}
If we let $g_j = {\rm e}^{-\omega_j t_0}\, p_j$ and define
\begin{equation}
C_{\ell\, \ell'} \equiv \sum_{{n(\ell)}\atop {n(\ell')}}  
                       g^*_j \, g_{j'}\, \langle a^\dagger_j a_{j'}\rangle
          \,, \label{eq:C}
\end{equation}
and 
\begin{equation}
D_{\ell\, \ell'} \equiv \bigl({\rm e}^{2 \Omega_{jj'}\, t_0} -1 \bigr) 
                                                    C_{\ell\, \ell'}
          \,, \label{eq:D}
\end{equation}
then we have that
\begin{equation}
S_m -  \tilde S_m = \sum_{jj'} (A_{jj'} - B_{jj'}) = \sum_{\ell, \ell' =0}^m 
          \bigl({\rm e}^{2 \Omega_{jj'}\, t_0}  -1 \bigr) C_{\ell\, \ell'}
                    = \sum_{\ell, \ell' =0}^m D_{\ell\, \ell'} \,.
\end{equation}
Our goal will be to prove that the right-hand side of this expression is
non-negative. 

As a prelude, let us note that any sum of the form 
$\sum_{\ell\, \ell'} C_{\ell\, \ell'}$, where $\ell$ and $\ell'$ range over 
the same set of values, is non-negative as a consequence of its being the
norm of a state vector. In particular, $C_{mm} \geq 0$. The quantity
$\sum_{\ell, \ell' =0}^m D_{\ell\, \ell'}$ is the sum of all of the elements
of an $(m+1)\times (m+1)$ matrix. Our plan is to prove the positivity
of this sum by working our way through the matrix, beginning in the 
lower right-hand corner. The element in this corner is
\begin{equation}
D_{mm} = \bigl({\rm e}^{2 \omega_m t_0} -1 \bigr) C_{mm} \geq 0 \,.
\end{equation}
Next consider the $2\times 2$ matrix formed by the four elements in this
corner. The sum over these elements is
\begin{eqnarray}
&&\!\!\!\!\!\!\!\!\!\!D_{m-1,m-1} +D_{m-1,m} +D_{m,m-1} +D_{m,m}  \nonumber \\
&=&\bigl({\rm e}^{2 \omega_{m-1} t_0} -1 \bigr)
                             \bigl(C_{m-1,m-1} +C_{m-1,m} +C_{m,m-1}\bigr)
+ \bigl({\rm e}^{2 \omega_m t_0} -1 \bigr)\, C_{mm} \nonumber \\
&\geq& \bigl({\rm e}^{2 \omega_{m-1} t_0} -1 \bigr)
\bigl(C_{m-1,m-1} +C_{m-1,m} +C_{m,m-1} +C_{mm}\bigr) \nonumber \\
&\geq& 0 \,.
\end{eqnarray}
Here we have used the facts that $C_{mm} \geq 0$,
$C_{m-1,m-1} +C_{m-1,m} +C_{m,m-1} +C_{mm} \geq 0$, and that 
${\rm e}^{2 \omega_m t_0} -1 \geq {\rm e}^{2 \omega_{m-1} t_0} -1$.

Now suppose that we have established that
\begin{equation}
\sum_{\ell, \ell' =L+1}^m D_{\ell\, \ell'} \geq 
\bigl({\rm e}^{2 \omega_{L+1} t_0} -1 \bigr)\,
                        \sum_{\ell, \ell' =L+1}^m C_{\ell\, \ell'}
  \geq 0 \,. \label{eq:sumL1}
\end{equation}
We wish to show that this implies that
\begin{equation}
\sum_{\ell, \ell' =L}^m D_{\ell\, \ell'} \geq 
\bigl({\rm e}^{2 \omega_L t_0} -1 \bigr)\,
                  \sum_{\ell, \ell' =L}^m C_{\ell\, \ell'}
  \geq 0 \,. \label{eq:sumL}
\end{equation}
The additional terms which are added in going from Eq.~(\ref{eq:sumL1})
to Eq.~(\ref{eq:sumL}) are those which lie in the same row as and to the
right of $D_{LL}$ and in the same column as and below $D_{LL}$, as 
illustrated below:
$$
\left[ \begin{array}{ccccccc}
D_{00}  & D_{01}    & \cdots      & {}          & {}     & {} & {} \\
D_{10}  & {}        & {}          & {}          & {}     & {} & {} \\
\vdots  & {}        & D_{LL}      & D_{L,L+1}   & \cdots & {} & D_{Lm}  \\
{}      & {}        & D_{L+1,L}   & D_{L+1,L+1} & \cdots & {} & D_{L+1,m}  \\
{}      & {}        & \vdots      & \vdots      & {}     & {} & \vdots \\
{}      & {}        & {}          & {}          & {}     & {} & {} \\
{}      & {}        & D_{mL}      & D_{m,L+1}   & \cdots & {} & D_{mm}  
\end{array}   \right] 
$$
Note
that $\omega_{\ell'}$ increases to the right and $\omega_{\ell}$ increases 
downward. Thus $\Omega_{\ell\, \ell'} = \omega_L$ for all of these 
elements. The sum of these terms is
\begin{eqnarray}
\!\!\!\!\!\!\!\!\!\!\!\!\!\!\!\!&\Delta D_L& \equiv D_{LL} +D_{L,L+1}+ 
                              \cdots +D_{Lm} +D_{L+1,L}+\cdots +D_{mL}
                                                       \nonumber \\
\!\!\!\!\!\!\!\!\!\!\!\!\!\!&=&\!\!\!\!\!\!\!\! 
                        \bigl({\rm e}^{2 \omega_L t_0} -1 \bigr)\,
\bigl(C_{LL} +C_{L,L+1}+ \cdots +C_{Lm} +C_{L+1,L}+\cdots +C_{mL}\bigr) \,.
\end{eqnarray}
Now we have that
\begin{eqnarray}
\sum_{\ell, \ell' =L}^m D_{\ell\, \ell'}  &=&   \Delta D_L 
 +\sum_{\ell, \ell' =L+1}^m D_{\ell\, \ell'} \nonumber \\
&\geq& \Delta D_L  + \bigl({\rm e}^{2\omega_{L+1} t_0} -1 \bigr)\,
            \sum_{\ell, \ell' =L+1}^m C_{\ell\, \ell'} \nonumber \\
&\geq& \Delta D_L  + \bigl({\rm e}^{2\omega_{L} t_0} -1 \bigr)\,
            \sum_{\ell, \ell' =L+1}^m C_{\ell\, \ell'} \nonumber \\
&=& \bigl({\rm e}^{2\omega_L t_0} -1 \bigr)\,
            \sum_{\ell, \ell' =L}^m C_{\ell\, \ell'} \nonumber \\
&\geq& 0\,.
\end{eqnarray}
 We have established that
Eq.~(\ref{eq:sumL}) holds for $L=m$ and for $L=m-1$. We have further
established that if it holds for one value of $L$, then it holds for the 
next smaller value of $L$. It now follows by induction that   
\begin{equation}
\sum_{\ell, \ell' =0}^m D_{\ell\, \ell'} \geq 0 \, ,
\end{equation}
and hence that 
\begin{equation}
S_m \geq  \tilde S_m   \,.
\end{equation}

Recall that the sums in both $S_m$ and $\tilde S_m$ are over a finite 
set of modes with maximum frequency $\omega_m$. However, we may now take 
the limit in which both the number of modes and $\omega_m$ become
infinite. So long as the resulting sums are convergent, which is the
case for the systems with which we are concerned, then we have that
\begin{equation}
\sum_{j j'}^\infty p^*_j \, p_{j'}\, \langle a^\dagger_j a_{j'}\rangle
           \, {\rm e}^{-|\omega_j -\omega_{j'}| t_0} \geq
  \sum_{j j'}^\infty p^*_j \, p_{j'}\, \langle a^\dagger_j a_{j'}\rangle
           \, {\rm e}^{-(\omega_j +\omega_{j'}) t_0} \,,  
 \label{eq:S_inf}
\end{equation}
 which is our final result.


\begin{thebibliography}{--}
\bibitem{MITCH/CBUMP} However, a similar situation occurs for a quantum 
                   inequality in the four-dimensional Einstein universe. See 
                   Ref.\cite{MITCH} for comparison.
\end{thebibliography}
\end{document}